\documentclass[a4paper,11pt]{article}
\usepackage{jheppub}
\usepackage[T1]{fontenc}
\usepackage{graphicx}
\usepackage{graphicx,booktabs,array}
\usepackage[latin1]{inputenc}
\usepackage{epsfig}
\usepackage{amsmath}
\usepackage{amsfonts}
\usepackage{float}
\usepackage{xcolor}
\usepackage{amssymb}
\usepackage[T1]{fontenc}
\usepackage{amssymb}
\usepackage{feynmp}
\usepackage[colorlinks=true
,urlcolor=blue
,citecolor=blue
,linkcolor=blue
,pagecolor=blue
,linktocpage=true
,pdfproducer=medialab
]{hyperref}
\usepackage{subfigure}
\usepackage{longtable}
\usepackage{multirow}
\usepackage{notoccite}
\usepackage[nodisplayskipstretch]{setspace}
\usepackage{placeins}
\graphicspath{{./Diagrams/}}
\def\be{\begin{equation}}
\def\ee{\end{equation}}
\def\bea{\begin{eqnarray}}
\def\eea{\end{eqnarray}}

\def\lsim{\raise0.3ex\hbox{$\;<$\kern-0.75em\raise-1.1ex\hbox{$\sim\;$}}}
\def\gsim{\raise0.3ex\hbox{$\;>$\kern-0.75em\raise-1.1ex\hbox{$\sim\;$}}}
\usepackage{epsfig}
\usepackage{xcolor}
\title{\textbf{Predictions for $\Lambda_b\to \Lambda_c\tau\bar\nu$ 
		in BLSSM with Inverse seesaw}}

\author[a,b]{\textbf{Dris Boubaa,}}
\author[c]{\textbf{ and Shaaban Khalil}}

\affiliation[b]{Department of Matter sciences, Faculty of Science and Technology,
	Abbes Laghrour University Khenchela, BP 1252 Road of Batna, Khenchela 40004, Algeria}
\affiliation[a]{Laboratoire de Physique des Particules et Physique Statistique, Ecole Normale Sup\'erieure-Kouba, 
	B.P. 92, 16050, Vieux-Kouba, Algiers, Algeria}
\affiliation[c]{Center for Fundamental Physics, Zewail City of
	Science and Technology, Sheikh Zayed,12588, Giza, Egypt}

\emailAdd{dris.boubaa@univ-khenchela.dz}
\emailAdd{skhalil@zewailcity.edu.eg}

\abstract{

The persistent deviations observed in semileptonic $B$ decays, in particular the lepton 
flavor universality ratios $\mathcal{R}(D^{(*)})$ and $\mathcal{R}(\Lambda_c)$, provide 
intriguing hints of physics beyond the Standard Model (SM). While current measurements 
remain limited by experimental uncertainties, their lower central values compared to SM 
expectations motivate further theoretical scrutiny. In this work we study these observables 
within the $B-L$ Supersymmetric Standard Model with an inverse seesaw (BLSSM-IS). We 
emphasize the role of penguin diagrams involving charginos, neutralinos, and right-handed 
sneutrinos, which induce flavor-dependent loop corrections to the effective $W\ell\nu$ vertex. 
These corrections can suppress the light-lepton decay rates relative to the $\tau$ mode, 
leading to a modest enhancement of $\mathcal{R}(D^{(*)})$ and, through the sum rule, 
$\mathcal{R}(\Lambda_c)$. We present updated numerical results illustrating the correlation 
between mesonic and baryonic observables, showing that the BLSSM-IS framework provides a 
natural and testable explanation of the current data. Our findings underline the importance 
of upcoming precision measurements at Belle II and the LHCb upgrade in clarifying the 
possible role of supersymmetry in lepton flavor universality violation.

		}		
		

\begin{document}
	\maketitle
	\flushbottom

\section{Introduction}
\label{sec:indtro}

The Standard Model (SM) of particle physics has achieved extraordinary success in accounting for a vast array of experimental observations. Yet, it fails to address several profound questions, including the particle nature of dark matter, the origin of neutrino masses and oscillations, the hierarchy and structure of fermion flavours, and the source of the observed matter-antimatter asymmetry, which is insufficiently explained by the SM mechanism of Charge-Parity (CP) violation. These open issues strongly suggest the existence of new physics (NP) beyond the SM, although direct evidence has so far remained elusive.

In this context, the study of B meson decays provides a unique laboratory for probing indirect effects of NP. Precision measurements in flavour physics, particularly in rare decays and CP-violating processes, have revealed intriguing tensions with SM expectations, such as possible hints of lepton flavour universality violation and anomalies in flavour-changing neutral currents. These observations make B physics one of the most promising frontiers in the search for NP.

Recent experimental measurements of semileptonic $B$-meson decays have suggested possible hints of lepton flavour universality (LFU) violation. If confirmed, such effects would constitute clear evidence for NP beyond the SM. In particular, the transitions $b \to c \ell \bar{\nu}_{\ell}$ ($\ell = e, \mu, \tau$) have been extensively investigated over the past decade. Continuous efforts by the \textsc{BaBar}, Belle, and LHCb collaborations have significantly improved the precision of these measurements, thereby providing an excellent testing ground for the SM and a sensitive probe of possible NP contributions. 

Semileptonic $B$ decays play a central role in examining the flavour sector of the SM, as they allow stringent tests of LFU and provide complementary information to rare $b \to s \ell^+ \ell^-$ transitions. To quantify potential deviations, it is convenient to construct LFU observables in terms of ratios of branching fractions between $\tau$ and light leptons ($\ell = e, \mu$). These ratios largely cancel hadronic uncertainties and thus serve as clean probes of possible NP effects. Explicitly, they are defined as
\begin{align}
{\cal R}(D) &=
\frac{{\rm BR}(B \to D \tau \bar\nu)}{{\rm BR}(B \to D \ell \bar\nu)}, 
& {\cal R}(D^{*}) &=
\frac{{\rm BR}(B \to D^{*} \tau \bar\nu)}{{\rm BR}(B \to D^{*} \ell \bar\nu)}, \\[6pt]
{\cal R}(J/\psi) &=
\frac{{\rm BR}(B_c \to J/\psi\, \tau \bar\nu)}{{\rm BR}(B_c \to J/\psi\, \ell \bar\nu)}, 
& {\cal R}(\Lambda_c) &=
\frac{{\rm BR}(\Lambda_b \to \Lambda_c \tau \bar\nu)}{{\rm BR}(\Lambda_b \to \Lambda_c \ell \bar\nu)}.
\end{align}

\begin{figure}[H]
	\centering
	\includegraphics[scale=0.45]{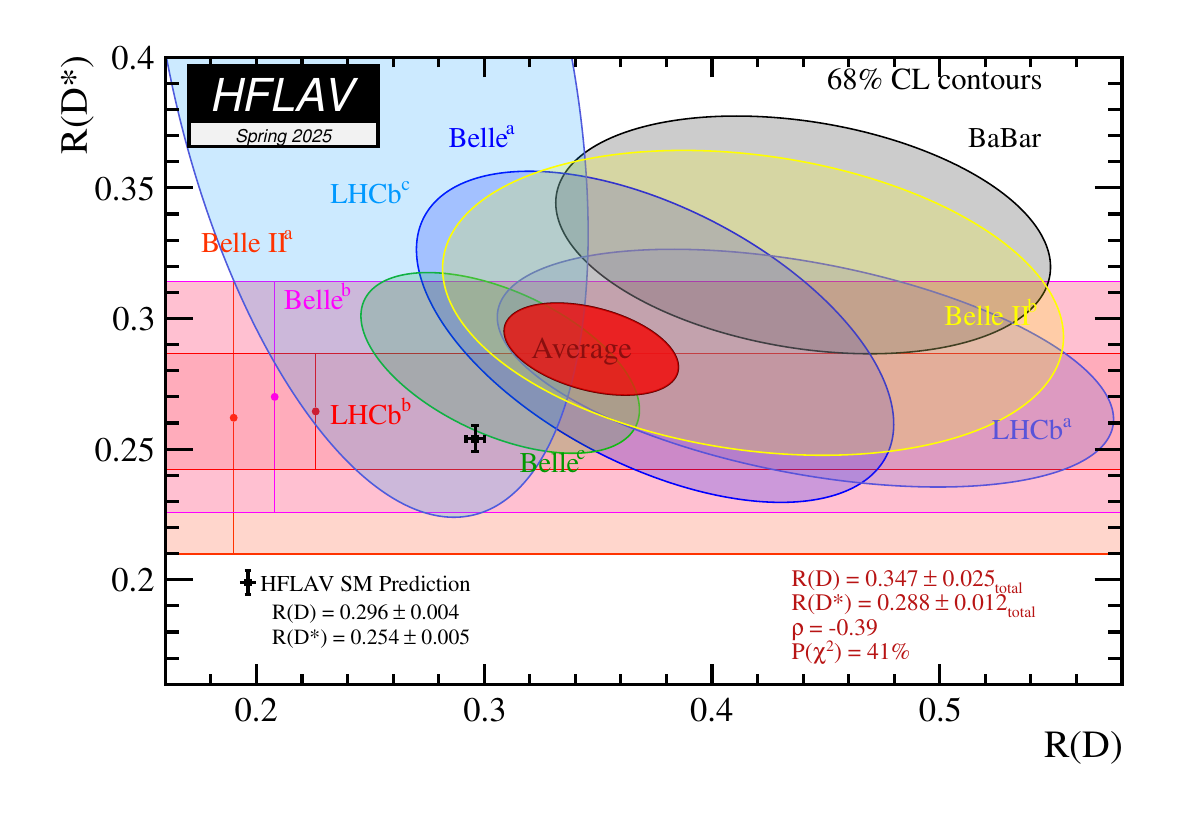}
	\caption{Experimental results for $\mathcal{R}(D)$ and $\mathcal{R}(D^*)$ reported by 
	LHCb, BaBar, and Belle between 2012 and 2024. The red ellipse represents the new world average, 
	while the black bars indicate the SM predictions. 
	Figure adapted from Ref.~\cite{HeavyFlavorAveragingGroupHFLAV:2024ctg}.}
	\label{figRD}
\end{figure}

The experimental measurements of $\mathcal{R}(D)$ and $\mathcal{R}(D^*)$ have been reported by the 
LHCb~\cite{LHCb:2015gmp,LHCb:2017smo,LHCb:2017rln}, 
BaBar~\cite{BaBar:2012obs,BaBar:2013mob}, 
and Belle~\cite{Belle:2019rba,Belle:2015qfa,Belle:2017ilt,Belle:2016dyj} collaborations 
between 2012 and 2024. These results are summarized in Fig.~\ref{figRD}, where the new world average 
is shown alongside the Standard Model predictions. 

The most recent averaged results for $\mathcal{R}(D)$ and $\mathcal{R}(D^*)$ have been reported by the 
Heavy Flavor Averaging Group (HFLAV)~\cite{HeavyFlavorAveragingGroupHFLAV:2024ctg} as
\begin{align}
\mathcal{R}(D) &= 0.347 \pm 0.025, \\
\mathcal{R}(D^*) &= 0.288 \pm 0.012.
\end{align}
These measurements deviate from the corresponding SM predictions by about $1.4\sigma$ for 
$\mathcal{R}(D)$ and $2.5\sigma$ for $\mathcal{R}(D^*)$, suggesting possible hints of LFU violation 
and the presence of NP contributions. The current SM expectations are given by 
Ref.~\cite{HeavyFlavorAveragingGroupHFLAV:2024ctg} as
\begin{align}
\mathcal{R}(D)_{\rm SM} &= 0.296 \pm 0.004, \\
\mathcal{R}(D^*)_{\rm SM} &= 0.254 \pm 0.005.
\end{align}

In contrast, the observable ${\cal R}(J/\psi)$ shows a larger apparent deviation from the SM, 
but its interpretation is complicated by sizable theoretical uncertainties in the hadronic 
transition form factors. Moreover, the available experimental measurements are less precise, 
which weakens its role as a stringent test of LFU 
(see Refs.~\cite{Tran:2018kuv,Watanabe:2017mip,Leljak:2019eyw,Fedele:2022iib}). 
For these reasons, we do not include ${\cal R}(J/\psi)$ in the present analysis.


 Another interesting observable, concerning the $\Lambda_b \to \Lambda_c \ell\bar\nu$ baryon decay recently observed by the LHCb collaboration, is the test of LFU in the ratio of semileptonic $b\to c \ell \bar\nu_\ell$ transitions \cite{LHCb:2022piu}:
\begin{equation}
\mathcal{R}(\Lambda_c) =  0.242 \;\pm\; 0.026_{\rm stat} \;\pm\; 0.040_{\rm syst} \;\pm\; 0.059_{\rm ext}.
\end{equation}
Combining the uncertainties in quadrature gives
\begin{equation}
\mathcal{R}(\Lambda_c)_{\rm exp} = 0.242 \pm 0.076.
\end{equation}
This should be compared with the SM prediction \cite{Li:2016pdv,Bernlochner:2018kxh,Datta:2017aue}
\begin{equation}
\mathcal{R}(\Lambda_c)_{\rm SM} = 0.324 \pm 0.004.
\end{equation}
The difference between the experimental and theoretical central values is
$\Delta \mathcal{R}(\Lambda_c) = -0.082$, corresponding to a deviation of about
\begin{equation}
\frac{|\Delta \mathcal{R}(\Lambda_c)|}{\sqrt{\sigma_{\rm exp}^2+\sigma_{\rm SM}^2}} 
\simeq 1.1\sigma.
\end{equation}
Thus, while the current result is statistically compatible with the SM, the lower central value may still point toward potential new physics effects. Improving the statistical precision in future measurements will be essential to clarify this possibility. 

Taken together, the observed deviations in $\mathcal{R}(D^{\ast})$ and $\mathcal{R}(\Lambda_c)$ relative to their SM predictions, albeit limited by present experimental uncertainties, constitute intriguing hints of possible LFU violation. Motivated by these anomalies, in this work we explore the correlation between $\mathcal{R}(D^{\ast})$ and $\mathcal{R}(\Lambda_c)$ within the $B\!-\!L$ extension of the Supersymmetric Standard Model  with an Inverse Seesaw mechanism (BLSSM-IS). In particular, we emphasize that new contributions from penguin diagrams involving charginos/neutralinos and right-handed sneutrinos can play a significant role, potentially reconciling the current tension between experimental measurements and SM expectations.

The paper is organized as follows. In Section~2, we summarize the effective Hamiltonian governing the $b \to c \ell \bar{\nu}_{\ell}$ transition for the processes $B \to D^{(*)} \ell \bar{\nu}$ and $\Lambda_b \to \Lambda_c \ell \bar{\nu}$. Section~3 provides a brief overview of the BLSSM-IS framework, with emphasis on the new particles that contribute to $b \to c \ell \bar{\nu}_{\ell}$ transitions. In Section~4, we present our new results on the BLSSM-IS contributions to $\mathcal{R}(D^{\ast})$ and $\mathcal{R}(\Lambda_c)$ and explore their possible correlation. Finally, conclusions and outlook are given in Section~5.


\section{Effective Hamiltonian of $b \to c$ Transitions}\label{sec42}

Semileptonic $b \to c \ell \bar{\nu}_{\ell}$ transitions, which govern the decays 
$B \to D^{(*)} \ell \bar{\nu}$ and $\Lambda_b \to \Lambda_c \ell \bar{\nu}$, can be described in terms of an effective Hamiltonian that incorporates the most general set of four-fermion operators:
\begin{equation}
\label{eq:effH}
\mathcal{H}_{\text{eff}} = \frac{4 G_F}{\sqrt{2}} V_{cb} \left[ \left(1 + C_{V_L}\right) \mathcal{O}_{V_L} + C_{V_R} \mathcal{O}_{V_R} + C_{S_L} \mathcal{O}_{S_L} + C_{S_R} \mathcal{O}_{S_R} + C_T \mathcal{O}_T \right] + \text{h.c.},
\end{equation}
where $G_F$ is the Fermi constant, $V_{cb}$ is the relevant CKM matrix element, and the effective operators are defined as  
\begin{align}
\mathcal{O}_{V_L} &= (\bar{c} \gamma^\mu P_L b)(\bar{\ell} \gamma_\mu P_L \nu_\ell), \\
\mathcal{O}_{V_R} &= (\bar{c} \gamma^\mu P_R b)(\bar{\ell} \gamma_\mu P_L \nu_\ell), \\
\mathcal{O}_{S_L} &= (\bar{c} P_L b)(\bar{\ell} P_L \nu_\ell), \\
\mathcal{O}_{S_R} &= (\bar{c} P_R b)(\bar{\ell} P_L \nu_\ell), \\
\mathcal{O}_T   &= (\bar{c} \sigma^{\mu\nu} P_L b)(\bar{\ell} \sigma_{\mu\nu} P_L \nu_\ell).
\end{align}
Here $P_{L,R} = \tfrac{1}{2}(1 \mp \gamma_5)$ are the chirality projection operators. The Wilson coefficients $C_i$ parameterize the short-distance physics and are generally model dependent. In the Standard Model, only the operator $\mathcal{O}_{V_L}$ contributes at tree level, corresponding to $C_{V_L} = 0$, while all other coefficients vanish.
The NP couplings $g_i$ are related to the Wilson coefficients by 
\begin{equation}
g_{i} = \frac{C_{i}^{\mathrm{NP}}}{C^{\mathrm{SM}}}, 
\qquad 
C^{\mathrm{SM}} = \frac{4 G_F}{\sqrt{2}} V_{cb}.
\end{equation}

Based on the above effective Hamiltonian, one can write the decay amplitude, $\mathcal{M}\equiv \langle D^{(\ast)}l\bar{\nu}_{l}|{\mathcal{H}}_{\mathrm{eff}}|\bar{B}\rangle$, of possible NP for the process $\bar{B}\rightarrow D^{(\ast)}l\bar{\nu}_{\ell}$
as 
\begin{equation}
\mathcal{M}= \mathcal{M}_{\mathrm SM}^{\lambda_{D^{(\ast)}},\lambda_{\ell}}+\mathcal{M}_{SL}^{\lambda_{D^{(\ast)}},\lambda_{\ell}}+\mathcal{M}_{SR}^{\lambda_{D^{(\ast)}},\lambda_{\ell}}+\mathcal{M}_{VL}^{\lambda_{D^{(\ast)}},\lambda_{\ell}}+\mathcal{M}_{VR}^{\lambda_{D^{(\ast)}},\lambda_{\ell}}+\mathcal{M}_{T}^{\lambda_{D^{(\ast)}},\lambda_{\ell}}
\end{equation}
where the SM amplitude is given by 
\begin{equation}
\mathcal{M}_{\mathrm SM}^{\lambda_{D^{(\ast)}},\lambda_{\ell}}\mathcal{=}\frac{G_{F}}{\sqrt{2}%
}V_{cb}\sum_{\lambda}\eta_{\lambda}H_{VL,\lambda}^{\lambda_{D^{(\ast)}}}L_{\lambda
}^{\lambda_{l}},
\end{equation}
and the NP ones are 
\begin{align}
\mathcal{M}_{S(L,R)}^{\lambda_{D^{(\ast)}},\lambda_{\ell}}  & \mathcal{=}%
\mathcal{-}\frac{G_{F}}{\sqrt{2}}V_{cb}g_{S(L,R)}H_{S(L,R)}^{\lambda_{D^{(\ast)}}%
}L^{\lambda_{\ell}},\\
\mathcal{M}_{V(L,R)}^{\lambda_{D^{(\ast)}},\lambda_{\ell}}  &  \mathcal{=}\frac{G_{F}%
}{\sqrt{2}}V_{cb}g_{V(L,R)}\sum_{\lambda}\eta_{\lambda}H_{V(L,R),\lambda
}^{\lambda_{D^{(\ast)}}}L_{\lambda}^{\lambda_{\ell}},\\
\mathcal{M}_{T}^{\lambda_{D^{(\ast)}},\lambda_{\ell}}    &\mathcal{=}\mathcal{-}%
\frac{G_{F}}{\sqrt{2}}V_{cb}g_{T}\sum_{\lambda,\lambda^{\prime}}\eta
_{\lambda^{\prime}}\eta_{\lambda}H_{\lambda\lambda^{\prime}}^{\lambda_{D^{(\ast)}}%
}L_{\lambda\lambda^{\prime}}^{\lambda_{\ell}}.
\end{align}
where the $D^{(\ast)}$-meson is either a spin-1 $D^{\ast}$-meson with
$\lambda_{D^{\ast}}=\pm,0$, or a spin-0 $D$-meson with $\lambda_{D}=s$. While $\lambda,\lambda^{\prime}=\pm,0$ or $s$ are the helicity of virtual vector bosons and $\lambda_{\ell}$ is the helicity of the lepton $\ell$. The summation is over the virtual vector boson helicities with the metric $\eta_{\pm}\ =\eta_{0}=-\eta_{s}=1$,$\ \ H$'s and $L$'s are the hadronic and leptonic amplitudes. An analogous behavior for the process $\Lambda_b \to\Lambda_c \ell\bar\nu$ with $\lambda_{D^{(\ast)}}\to\lambda_{\Lambda_c}=\pm\frac{1}{2}$   (for more details, see Refs.~\cite{Datta:2017aue,Tanaka:2012nw}). 

 On the other hand, the differential decay rate for the process $\bar{B}\rightarrow D^{(\ast)}l\bar{\nu}_{\ell}$ is defined 
		\be
		\frac{d\Gamma}{dq^{2}d\cos\theta_{\ell}}=\frac{\sqrt{Q_{+}Q_{-}}}{256\pi^{3}%
			m_{B}^{3}}\Big(1-\frac{m_{\ell}^{2}}{q^{2}}\Big)|\mathcal{M}(q^2,\cos\theta_{\ell})|^{2},
		\label{diff_decayrate}
		\ee
		where $Q_{\pm}=(m_{B}\pm m_{D^{(\ast)}})^{2}-q^{2}$ with $q^{\mu}=p_{B}^{\mu}-p_{D^{(\ast)}}^{\mu}=p_{\ell}^{\mu}+p_{\nu_{\ell}}^{\mu}$,  $q^{2}$ varies in the range $m_{\ell}^{2}\leq q^{2}$
		$\leq(m_{B}-m_{D^{(\ast)}})^{2}$ and $-1\leq\cos\theta_{\ell}\leq1$. For the process $\Lambda_b \to\Lambda_c \ell\bar\nu$ we make the exchanges $(m_{B},m_{D^{(\ast)}})\to (m_{\Lambda_b},m_{\Lambda_b})$ and $(p_{B}^{\mu},p_{D^{(\ast)}}^{\mu})\to(p_{\Lambda_b}^{\mu},p_{\Lambda_c}^{\mu})$.

B meson and $ \Lambda_b$ decays play an important role for well understanding of the ${\cal R}(D^{(\ast)})$ and ${\cal R}(\Lambda_c)$ anomalies through the quark level transition $b\to c \ell \bar\nu$ decay. This transition is described by the effective Hamiltonian of Eq.~(\ref{eq:effH}) including the effects of NP which may be considered through the Wilson coefficients ratios $g_i$.
 
The observables  ${\cal R}(D^{(\ast)})$ and ${\cal R}(\Lambda_c)$ can be defined as

\begin{align}
\mathcal{R}(D)&=\frac{{\rm BR}(B\to D \tau \bar\nu)}{{\rm BR}(B\to D \ell \bar\nu)}=\frac{\Gamma({B}\rightarrow D\tau\nu_{\tau})}{\Gamma({B}\rightarrow D \ell\nu_{\ell})}=\frac{\int_{m_\tau^2}^{(m_B-m_D)^2}\frac{d\Gamma_{\tau}^D}{dq^2}dq^2}{\int_{m_\ell^2}^{(m_B-m_D)^2}\frac{d\Gamma^D_l}{dq^2}dq^2},\\
\mathcal{R}(D^{\ast})&=
\frac{{\rm BR}(B\to D^{*} \tau \bar\nu)}{{\rm BR}(B\to D^{*} \ell \bar\nu)}=\frac{\Gamma(\bar{B}\rightarrow D^{\ast}\tau\nu_{\tau})}{\Gamma(\bar{B}\rightarrow D^{\ast}\ell\nu_{\ell})}=\frac{\int_{m_\tau^2}^{(m_B-m_D^{\ast})^2}\frac{d\Gamma_{\tau}^{D^{\ast}}}{dq^2}dq^2}{\int_{m_l^2}^{(m_B-m_D^{\ast})^2}\frac{d\Gamma^{D^{\ast}}_\ell}{dq^2}dq^2},\\
\mathcal{R}(\Lambda_c)&=\frac{{\rm BR}(\Lambda_b\to \Lambda_c \tau \bar\nu)}{{\rm BR}(\Lambda_b\to \Lambda_c \ell \bar\nu)}=\frac{\Gamma(\Lambda_b\rightarrow \Lambda_c\tau\nu_{\tau})}{\Gamma(\Lambda_b\rightarrow \Lambda_c \ell\nu_{\ell})}=\frac{\int_{m_\tau^2}^{(m_{\Lambda_b}-m_{\Lambda_c})^2}\frac{d\Gamma_{\tau}^{\Lambda_c}}{dq^2}dq^2}{\int_{m_\ell^2}^{(m_{\Lambda_b}-m_{\Lambda_c})^2}\frac{d\Gamma^{{\Lambda_c}}_l}{dq^2}dq^2}
\end{align}
 The updated numerical formulae of $\mathcal{R}(D^{(\ast)})$ and $\mathcal{R}(\Lambda_c)$ are \cite{Iguro:2022yzr,Fedele:2022iib,Detmold:2015aaa,Becirevic:2018afm}
 \begin{align}
 \label{eq:RD}
 \frac{\mathcal{R}({D})}{\mathcal{R}(D)^\textrm{SM}} =
 & ~|1+g_{V_L}+g_{V_R}|^2  + 1.01|g_{S_L}+g_{S_R}|^2 + 0.84|g_{T}|^2  \nonumber \\[-0.5em]
 & + 1.49\textrm{Re}[(1+g_{V_L}+g_{V_R})(g_{S_L}^*+g_{S_R}^*)]\nonumber\\  &+ 1.08\textrm{Re}[(1+g_{V_L}+g_{V_R})g_{T}^*] \,, 
 \\[0.5em]
 \label{eq:RDs}
 \frac{ \mathcal{R}({D^{\ast}})}{\mathcal{R}(D^\ast)^\textrm{SM}} =
 & ~|1+g_{V_L}|^2 + |g_{V_R}|^2  + 0.04|g_{S_L}-g_{S_R}|^2 + 16.0|g_{T}|^2 \nonumber \\[-0.5em]
 & -1.83\textrm{Re}[(1+g_{V_L})g_{V_R}^*]  - 0.11\textrm{Re}[(1+g_{V_L}-g_{V_R})(g_{S_L}^*-g_{S_R}^*)] \nonumber \\ 
 & -5.17\textrm{Re}[(1+g_{V_L})g_{T}^*] + 6.60\textrm{Re}[g_{V_R}g_{T}^*] \,, 
 \\[0.5em]
 \label{eq:RLambda}
 \frac{\mathcal{R}(\Lambda_c)}{\mathcal{R}(\Lambda_c)^{\textrm{SM}}}&=  \left|1+g_{V_L}\right|^2
 + 0.13\,\textrm{Re}\!\left[\,(1+g_{V_L})\,g_{V_R}^{\ast}\right]  + 0.04\,|g_{V_R}|^2
+ 0.50\,\textrm{Re}\!\left[(1+g_{V_L})\,g_{S_R}^{\ast}\right] \nonumber\\
& + 0.33 \,\textrm{Re}\left[ \left(1 +g_{V_L}\right) g_{S_L}^{\ast} 
 \right] + 0.52~\textrm{Re}\left( g_{S_L} g_{S_R}^{\ast}  \right)+ 0.32\left(|g_{S_L}|^2 + |g_{S_R}|^2\right)\nonumber\\
 &-3.11 \,\textrm{Re}\left[ \left(1 +g_{V_L}\right)  g_T^{\ast}\right]
 + 10.4 |g_T|^2.
 \end{align}
\begin{figure*}[h]
	\centering
	\subfigure{\includegraphics[scale=0.4]{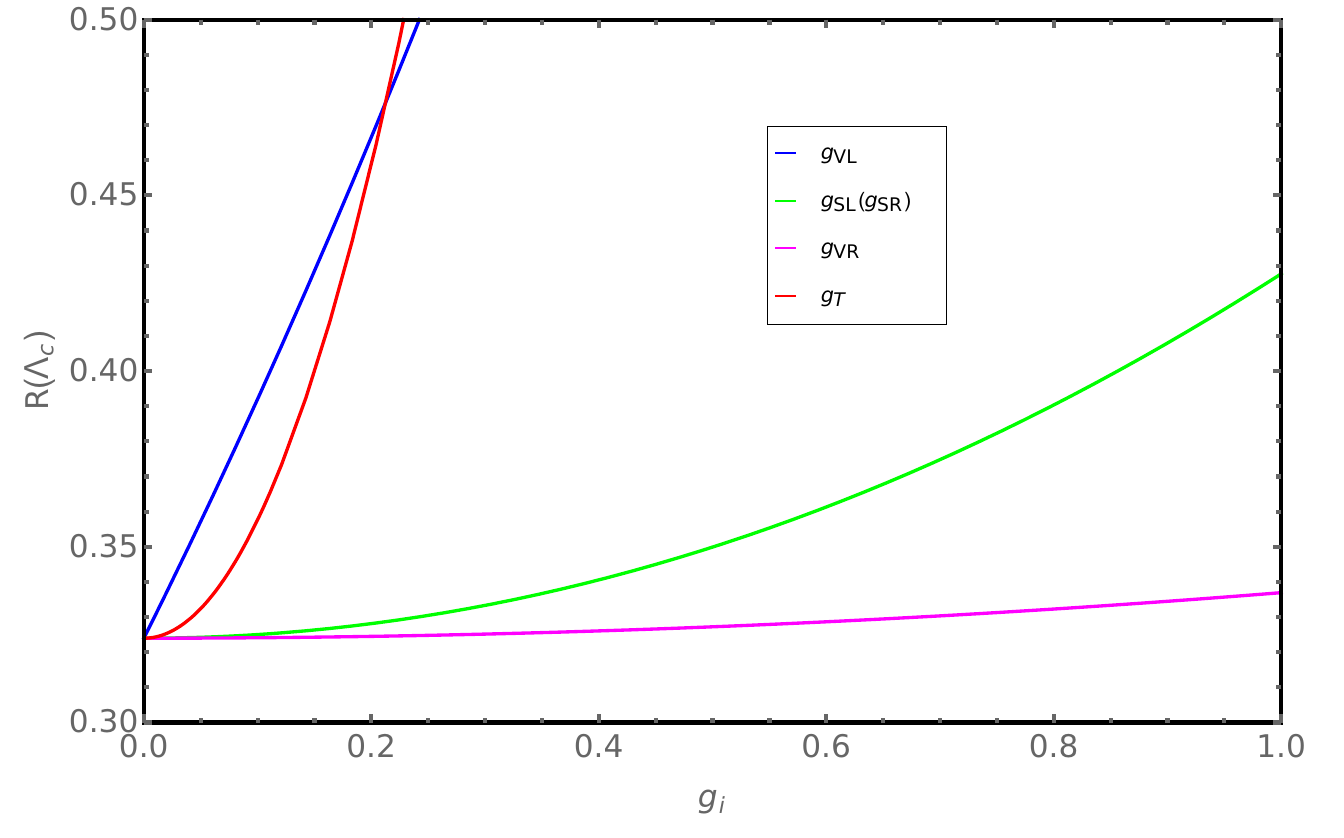}}
	\caption{$\mathcal{R}(\Lambda_c)$ as function of the individual Wilson coefficients $g_i$.}
	\label{fig2}
\end{figure*}
Because the same NP couplings enter mesonic and baryonic modes, one finds a powerful correlation \cite{Blanke:2018yud,Blanke:2019qrx,Fedele:2022iib}:
\begin{equation}
\frac{\mathcal{R}(\Lambda_c)}{\mathcal{R}(\Lambda_c)_{\rm SM}}
=0.280\,\frac{\mathcal{R}(D)}{\mathcal{R}(D)_{\rm SM}}
+0.720\,\frac{\mathcal{R}(D^{*})}{\mathcal{R}(D^{*})_{\rm SM}}
+\delta_{\Lambda_c},
\label{eq:sumrule}
\end{equation}
with a small correction term
\be
\delta_{\Lambda_c} = 
\textrm{Re}\Big[(1+g_{V_L})\big(0.314\,g_T^{*}-0.003\,g_{S_R}^{*}\big)\Big]
+0.014 |g_{S_L}|^2
+0.004 \textrm{Re}(g_{S_L} g_{S_R}^{\ast})
+|g_{S_R}|^2
-1.3 |g_T|^2 .
\label{eq:delta}
\ee

It follows that if  $\mathcal{R}(D)=\mathcal{R}(D)_{\rm SM}$ and $\mathcal{R}(D^{*})=\mathcal{R}(D^{*})_{\rm SM}$,
then the effective new-physics contributions to the $b \to c \tau \nu$ transition must vanish within the current experimental precision. This implies that all NP-induced Wilson coefficients are negligible, and consequently
\[
\mathcal{R}(\Lambda_c)=\mathcal{R}(\Lambda_c)_{\rm SM}.
\]
This conclusion is robust unless one allows for highly fine-tuned cancellations among different operators. A few remarks are in order:
\begin{itemize}
\item If NP is dominated by $g_{V_L}$ or $g_{V_R}$, then $\delta_{\Lambda_c}\simeq 0$, and the sum rule~\eqref{eq:sumrule} holds to very high accuracy.  

\item For purely scalar NP with only $g_{S_L}\neq 0$, the relation remains valid with a small correction, 
\[
\delta_{\Lambda_c} = 0.014\,|g_{S_L}|^2,
\]
which is negligible in practice (below the $0.1\%$ level for $|g_{S_L}|\lesssim 0.3$).  

\item For sizeable scalar or tensor contributions, $\delta_{\Lambda_c}$ can become non-negligible and must be taken into account explicitly.  
\end{itemize}
In Fig.~\ref{fig2}, we display the variation of $\mathcal{R}(\Lambda_c)$ with respect to the individual Wilson coefficients $g_i$. A nonzero left-handed vector coupling $g_{VL}$ produces a nearly linear enhancement of $\mathcal{R}(\Lambda_c)$, reflecting its direct interference with the SM contribution. On the other hand, the right-handed vector, scalar, and pseudoscalar couplings yield only mild deviations from the SM prediction, with their effects being relatively suppressed. In contrast, tensor interactions exhibit a quadratic dependence, leading to a pronounced enhancement of $\mathcal{R}(\Lambda_c)$ even for moderate values of $g_T$.

The ratios of leptonic $W$-boson decay widths provide sensitive probes of lepton flavour universality (LFU) and possible new-physics effects relevant for $R_{\Lambda_c}$, $R_{D}$, and $R_{D^*}$. Experimentally, one finds~\cite{ParticleDataGroup:2024cfk}  
\bea 
W_{\tau e} &\equiv& \frac{\Gamma(W\rightarrow\tau\nu)}{\Gamma(W\rightarrow e\nu)}  = 1.015 \pm 0.020, \\
W_{\tau \mu} &\equiv& \frac{\Gamma(W\rightarrow\tau\nu)}{\Gamma(W\rightarrow\mu\nu)}  = 1.002 \pm 0.020.
\label{eq:Wleps}
\eea  

The SUSY contributions to these decays can be parametrised as  
\begin{equation}%
\Gamma(W\rightarrow \ell\nu)=\frac{G_{F}M_{W^\pm}^{3}}{6\sqrt{2}\pi}
\left(1-\frac{m_{\ell}^{2}}{M_{W^\pm}^{2}}\right)^{2}
\left(1+\frac{1}{2}\frac{m_{\ell}^{2}}{M_{W^\pm}^{2}}\right)
\big|1+g_{VL}^{\prime}\big|^{2},
\end{equation}
where $g_{VL}^{\prime}=C^{\rm SUSY}(W\rightarrow \ell\nu)/C^{\rm SM}(W\rightarrow \ell\nu)$ and $C^{\rm SM}(W\rightarrow \ell\nu)=g/\sqrt{2}$.  

Similarly, the ratios of leptonic $\tau$ decays also provide precise tests of LFU. Defining  
\be\label{LN}
\tau_{\mu e}\equiv\frac{\Gamma(\tau \to \mu\nu_{\tau}\nu_{\mu})}{\Gamma(\tau \to e\nu_{\tau}\nu_{e})}
=0.9726\,\frac{|1+g_{VL}^{\mu}|^{2}}{|1+g_{VL}^{e}|^{2}},
\ee  
one obtains the experimental value $\tau_{\mu e}=0.979\pm 0.004$~\cite{ParticleDataGroup:2024cfk}.  
Here $g_{VL}^{\ell}=C^{\rm SUSY}(\tau \to \nu_{\tau}\ell\nu_{\ell})/C^{\rm SM}(\tau \to \nu_{\tau}\ell\nu_{\ell})$, with $C^{\rm SM}(\tau \to \nu_{\tau}\ell\nu_{\ell})=2\sqrt{2}G_{F}$. The high precision of these measurements makes them a powerful tool for constraining LFU violation in SUSY scenarios.  

Further constraints arise from rare $B$-meson decays, such as $B_{s}\rightarrow X_{s}\gamma$ and $B_{s}\rightarrow \mu^{+}\mu^{-}$~\cite{HeavyFlavorAveragingGroupHFLAV:2024ctg}. In addition, stringent bounds are provided by searches for charged lepton-flavour-violating (LFV) processes, notably $l_{i}\rightarrow l_{j}\gamma$~\cite{MEGII:2021fah,BaBar:2009hkt}. Direct searches also impose robust lower limits on the masses of SUSY particles~\cite{ParticleDataGroup:2024cfk,ATLAS:2012yve,CMS:2012qbp,ATLAS:2017mjy,ATLAS:2019npw}.  
Electroweak precision data provide complementary information. In particular, the oblique parameters $S$, $T$, and $U$ constrain potential deviations from the Standard Model. A closely related quantity is the $\rho$ parameter, defined as~\cite{ParticleDataGroup:2014cgo}  
\begin{equation}
\rho - 1 = \frac{1}{1-\hat{\alpha}(M_Z)T} \simeq \hat{\alpha}(M_Z)T,
\end{equation}
where $\hat{\alpha}(M_Z)$ denotes the renormalised electromagnetic coupling constant at the $M_Z$ scale.  
Taken together, these observables provide stringent bounds on new-physics effects, significantly constraining possible SUSY contributions to LFU violation and rare processes.

\section{BLSSM-IS and Contributions to $\mathcal{R}(\Lambda_c)$ and $\mathcal{R}(D^{(\ast)})$}
\subsection{BLSSM-IS}
Although the absence of direct experimental evidence, supersymmetry (SUSY) remains one of the most promising frameworks for a unified theory beyond the Standard Model (SM). The simplest realization of such an extension is the Minimal Supersymmetric Standard Model (MSSM), based on the gauge group 
\[
SU(3)_C \times SU(2)_L \times U(1)_Y.
\]
However, the MSSM admits interactions that violate baryon and lepton number conservation. Since the global $B-L$ symmetry is preserved in the SM and tested to high precision, an additional discrete, $R$-parity, is typically imposed to forbid such terms. A more natural way to realize $R$-parity is by gauging the $U(1)_{B-L}$ symmetry, which is spontaneously broken above the electroweak scale. This leads to the $B-L$ Supersymmetric Model (BLSSM), characterized by extended gauge group and richer particle content:
\[
SU(3)_C \times SU(2)_L \times U(1)_Y \times U(1)_{B-L}.
\]

In the BLSSM with Inverse Seesaw (BLSSM-IS), the MSSM particle spectrum is extended as follows \cite{Khalil:2010iu}:  
\begin{itemize}
    \item Two SM-singlet chiral Higgs superfields, $\hat{\chi}_{1,2}$, whose scalar components acquire Vacuum Expectation Values (VEVs) that spontaneously break $U(1)_{B-L}$. The field $\hat{\chi}_2$ is also required for anomaly cancellation.  
    \item Three generations of SM-singlet chiral superfields, $\hat{\nu}^c_i$, $\hat{s}_{1_i}$, and $\hat{s}_{2_i}$ ($i=1,2,3$), introduced to implement the Inverse Seesaw (IS) mechanism and ensure anomaly cancellation.  
\end{itemize}
BLSSM-IS and their $B-L$ charges are summarized in Table~\ref{tab:BLSSMfields}.  
\begin{table}[h!]
\centering
\begin{tabular}{|c|c|}
\hline
\textbf{Superfield} & \textbf{$B-L$ Charge} \\
\hline
$\hat{\chi}_1$ & $+1$ \\
$\hat{\chi}_2$ & $-1$ \\
$\hat{\nu}^c_i$ & $-1$ \\
$\hat{s}_{1_i}$ & $+2$ \\
$\hat{s}_{2_i}$ & $-2$ \\
\hline
\end{tabular}
\caption{Additional superfields of the BLSSM-IS beyond the MSSM spectrum.}
\label{tab:BLSSMfields}
\end{table}
The inclusion of these new fields allows for a consistent anomaly-free construction, spontaneous breaking of the $U(1)_{B-L}$ symmetry, and the implementation of the Inverse Seesaw mechanism. Consequently, the BLSSM-IS provides a natural framework for generating small neutrino masses while keeping the associated new physics at the TeV scale.  
The superpotential of the BLSSM-IS model is then given by
\begin{align}
W = Y_u \hat{Q} \hat{H}_2 \hat{U}^c + Y_d \hat{Q} \hat{H}_1 \hat{D}^c + Y_e \hat{L} \hat{H}_1 \hat{E}^c + Y_\nu \hat{L} \hat{H}_2 \hat{\nu}^c + Y_S \hat{\nu}^c \hat{\chi}_1 \hat{S}_2   + \mu \hat{H}_1 \hat{H}_2 + \mu' \hat{\chi}_1 \hat{\chi}_2.
\label{superpotential}
\end{align}
In this superpotential, the terms $Y_\nu \hat{L} \hat{H}_2 \hat{\nu}^c$ and $Y_S \hat{\nu}^c \hat{\chi}_1 \hat{S}_2$ are responsible for generating neutrino masses via the inverse seesaw mechanism, with the smallness of the light neutrino masses controlled by the mass scale of the singlet fermions $S_2$ and the vacuum expectation value of $\hat{\chi}_1$.

The $B-L$ symmetry is radiatively broken via the VEVs of $\chi_{1,2}$:  $\langle{\rm Re} \,\chi^0_i\rangle=\frac{v'_i}{\sqrt{2}}, \quad (i=1,2)$, 
while the electroweak symmetry is broken by the VEVs of $H_{u,d}$: $ \langle{\rm Re}\,H_{u,d}^0\rangle=\frac{v_{u,d}}{\sqrt{2}}, \qquad v=\sqrt{v_u^2+v_d^2}=246~{\rm GeV}, \quad v'=\sqrt{v'^2_1+v'^2_2}\sim {\cal O}(1)~{\rm TeV}$. The ratios are defined as $\tan\beta=v_u/v_d$ and $\tan\beta'=v'_1/v'_2$ \cite{Khalil:2007dr}.  
In the context of the BLSSM-IS, we now focus on the particle spectrum most relevant to the transition $b \to c \ell \bar{\nu}$. The key states can be summarized as follows:
\begin{itemize}
    \item \textbf{Lightest right-handed sneutrino, $\tilde{\nu}_1$:}  
     This is the lightest CP-even/CP-odd combination arising from the fields the mixing of the right-handed sneutrino $\tilde{\nu}_R$ with the singlet $\tilde{S}_2$. Its mass can lie in the range $400$--$1400$~GeV \cite{Elsayed:2011de,Khalil:2015naa,Khalil:2016lgy}, and it can contribute to semileptonic $B$ decays via sneutrino-mediated interactions.  
    
    \item \textbf{Lightest neutralino, $\tilde{\chi}^0_1$:}  
    The neutralino sector of the BLSSM-IS is enlarged to include the gaugino $\tilde{B}'$ and the Higgsinos $\tilde{\chi}_{1,2}$ in addition to the MSSM states. The lightest neutralino may thus be bino-, Higgsino-, $B'$ino-, or $\chi$-like:
 \begin{equation}
        \tilde{\chi}^0_1 = V_{11} \tilde{B} + V_{12} \tilde{W}^3 + V_{13} \tilde{H}_1^0 + V_{14} \tilde{H}_2^0 + V_{15} \tilde{B}' + V_{16} \tilde{\chi}_1 + V_{17} \tilde{\chi}_2,
    \end{equation}
with typical mass of ${\cal O}(100~\text{GeV})$. 

    \item \textbf{Lightest chargino, $\tilde{\chi}^\pm_1$:}  
    The chargino sector remains MSSM-like in the BLSSM-IS, consisting of the charged wino and Higgsino states. The lightest chargino typically has a mass at the electroweak/TeV scale and can play a role in charged-current processes.  

    \item \textbf{Right-handed neutrinos, $\nu_{R_i}$ ($i=1,\dots,6$):}  
    The inverse seesaw mechanism introduces six additional singlet fermions, which pair up with the right-handed neutrinos. Their mass spectrum is typically at the TeV scale, and their presence is crucial for explaining the smallness of the observed light neutrino masses.
\end{itemize}

\subsection{Contributions to $\mathcal{R}(\Lambda_c)$ and $\mathcal{R}(D^{(\ast)})$}

\begin{figure*}[!]
	\centering
	\subfigure{\includegraphics[scale=0.65]{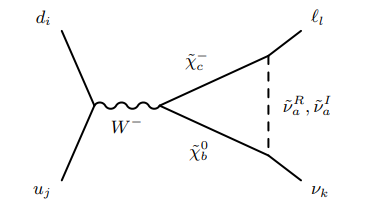}}~~~
	\subfigure{\includegraphics[scale=0.75]{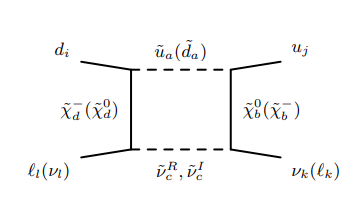}}\\
	\caption{Penguin and Box diagrams in the BLSSM-IS contributing to the $b \to c\ell \bar\nu_{\ell}$ transition.}
	\label{fig3}
\end{figure*}

The BLSSM-IS contributions to the ratios $\mathcal{R}(\Lambda_c)$ and $\mathcal{R}(D^{(\ast)})$ primarily arise from penguin-type corrections to the vertex $W^\pm \ell \nu_\ell$ $(\ell = e, \mu, \tau)$, mediated by the exchange of charginos and neutralinos together with right-handed sneutrinos, as illustrated in Fig.~\ref{fig3}. A key feature of the BLSSM-IS is the large neutrino Yukawa coupling, $Y_\nu \sim \mathcal{O}(1)$, between the right-handed sneutrino and the chargino/neutralino sector and the charged lepton/neutrino states. This sizable coupling significantly enhances the loop corrections to the $b \to c \tau \nu$ transition.

In addition to these penguin diagrams, one should note that box diagrams involving charginos, neutralinos, and right-handed sneutrinos can also contribute to the effective four-fermion operators relevant for $b \to c \tau \nu$. Although these box-induced effects are typically subleading compared to the penguin contributions, they may become non-negligible in certain regions of the parameter space, particularly if the sneutrino and chargino states are relatively light. Therefore, while the dominant new-physics effects in $\mathcal{R}(\Lambda_c)$ and $\mathcal{R}(D^{(\ast)})$ stem from penguin corrections, box contributions provide an additional source of deviation from the SM and MSSM predictions in the BLSSM-IS framework.


 The relevant Wilson coefficient is given by
\begin{align}        
C_{VL}^{{\textrm{Penguin}}}&=  \frac{G_{\tilde{\chi_b}^0 \nu_k \tilde\nu^R_a}^L G_{\bar{l_l}\tilde{\chi}_c^- \tilde\nu_a^R }^R G_{\bar{u_j}d_i W^-}^L}{16\pi^2m^2_{W^-}}  \Big[-G_{\tilde{\chi}_c^-\tilde{\chi}_b^0 W^- }^L m_{\tilde{\chi}^0_{{b}}} m_{\tilde{\chi}^-_{{c}}}
C_0(m^2_{\tilde{\chi}^-_{{c}}}, m^2_{\tilde{\chi}^0_{{b}}}, m^2_{\tilde\nu_a^R})\nonumber\\
& + G_{\tilde{\chi}_c^-\tilde{\chi}_b^0 W^-}^R \big[B_0(m^2_{\tilde{\chi}^0_{{b}}}, m^2_{\tilde{\chi}^-_{{c}}}) 
- 2 C_{00}(m^2_{\tilde{\chi}^-_{{c}}}, m^2_{\tilde{\chi}^0_{{b}}}, m^2_{\tilde\nu_a^R}) \!+\! m^2_{\tilde\nu_a^R}C_0(m^2_{\tilde{\chi}^-_{{c}}}, m^2_{\tilde{\chi}^0_{{b}}}, m^2_{\tilde\nu_a^R})\!\big]\!\Big]\!,~~~
\\
C_{VL}^{\textrm{Box}}= &\frac{1}{32\pi^2} G^{L}_{\tilde{\chi}^-_d d_i \tilde{u}^*_a} G^{R}_{\tilde{\chi}^0_b {u}_j\tilde{u}_a} G^{L}_{\tilde{\chi}^0_d \nu_k \tilde\nu^R_c} G^{R}_{{\ell}_l\tilde{\chi}^-_b \tilde\nu^R_c} m_{\tilde{\chi}^0_{{b}}} m_{\tilde{\chi}^-_{{d}}}D_{0}(m^2_{\tilde{\chi}^0_{{d}}}, m^2_{\tilde{\chi}^-_{{b}}}, m^2_{\tilde{u}_{{a}}}, m^2_{\nu^R_{{c}}}),
\\
C_{VR}^{\textrm{Box}}= &\frac{1}{16\pi^2} G^{R}_{\tilde{\chi}^-_d d_i \tilde{u}^*_a} G^{L}_{\tilde{\chi}^0_b {u}_j\tilde{u}_a} G^{R}_{\tilde{\chi}^0_d \nu_k \tilde\nu^R_c} G^{L}_{{\ell}_l\tilde{\chi}^-_b \tilde\nu^R_c} D_{27}(m^2_{\tilde{\chi}^0_{{d}}}, m^2_{\tilde{\chi}^-_{{b}}}, m^2_{\tilde{u}_{{a}}}, m^2_{\nu^R_{{c}}}),
\\
C_{SL}^{\textrm{Box}}= &\frac{1}{32\pi^2} G^{L}_{\tilde{\chi}^-_d d_i \tilde{u}^*_a} G^{L}_{\tilde{\chi}^0_b {u}_j\tilde{u}_a} G^{R}_{\tilde{\chi}^0_d \nu_k \tilde\nu^R_c} G^{R}_{{\ell}_l\tilde{\chi}^-_b \tilde\nu^R_c} m_{\tilde{\chi}^0_{{b}}} m_{\tilde{\chi}^-_{{d}}}D_{0}(m^2_{\tilde{\chi}^0_{{d}}}, m^2_{\tilde{\chi}^-_{{b}}}, m^2_{\tilde{u}_{{a}}}, m^2_{\nu^R_{{c}}}),
\\
C_{SR}^{\textrm{Box}}= &\frac{1}{8\pi^2} G^{R}_{\tilde{\chi}^-_d d_i \tilde{u}^*_a} G^{R}_{\tilde{\chi}^0_b {u}_j\tilde{u}_a} G^{R}_{\tilde{\chi}^0_d \nu_k \tilde\nu^R_c} G^{R}_{{\ell}_l\tilde{\chi}^-_b \tilde\nu^R_c} D_{27}(m^2_{\tilde{\chi}^0_{{d}}}, m^2_{\tilde{\chi}^-_{{b}}}, m^2_{\tilde{u}_{{a}}}, m^2_{\nu^R_{{c}}}),
\\
C_{T}^{\textrm{Box}}= &\frac{1}{128\pi^2} G^{L}_{\tilde{\chi}^-_d d_i \tilde{u}^*_a} G^{L}_{\tilde{\chi}^0_b {u}_j\tilde{u}_a} G^{R}_{\tilde{\chi}^0_d \nu_k \tilde\nu^R_c} G^{R}_{{\ell}_l\tilde{\chi}^-_b \tilde\nu^R_c} m_{\tilde{\chi}^0_{{b}}} m_{\tilde{\chi}^-_{{d}}}D_{0}(m^2_{\tilde{\chi}^0_{{d}}}, m^2_{\tilde{\chi}^-_{{b}}}, m^2_{\tilde{u}_{{a}}}, m^2_{\nu^R_{{c}}}),
\end{align}
where $\tilde{\nu}^R$ denotes the CP-even right-handed sneutrino, while the substitution $R \to I$ corresponds to the CP-odd right-handed sneutrino. The symbol $G$ represents the couplings among the particles appearing in the loops of Fig.~\ref{fig3}. The loop functions $C_{0}(x,y,z)$, $B_{0}(x,y)$, $C_{00}(x,y,z)$, $D_{0}(x,y,z,w)$, and $D_{27}(x,y,z,w)$ are defined in Ref.~\cite{Abada:2014kba}. Analogous contributions are induced by $\tilde{d}$ exchange, obtained through the substitutions $\tilde{\chi}^- \to \tilde{\chi}^0$ and $\tilde{u} \to \tilde{d}$.

The loop functions associated with the box diagrams, $D_{0}$ and $D_{27}$, are negligible, being suppressed by more than four orders of magnitude relative to the penguin loop function $C_{0}$. Therefore, in the following analysis we focus exclusively on the penguin contributions, while safely neglecting those from the box diagrams. Moreover, in order to enhance the penguin loop function, it is preferable that the masses of the particles running in the penguin diagrams (chargino, neutralino, and right-handed sneutrino) remain moderately light and nearly degenerate, as emphasized in Ref.~\cite{Boubaa:2022xsk,Boubaa:2020ksf,Boubaa:2016mgn}. This framework will be adopted in the present analysis.

\begin{figure*}
	\centering
	\subfigure{\includegraphics[scale=0.4]{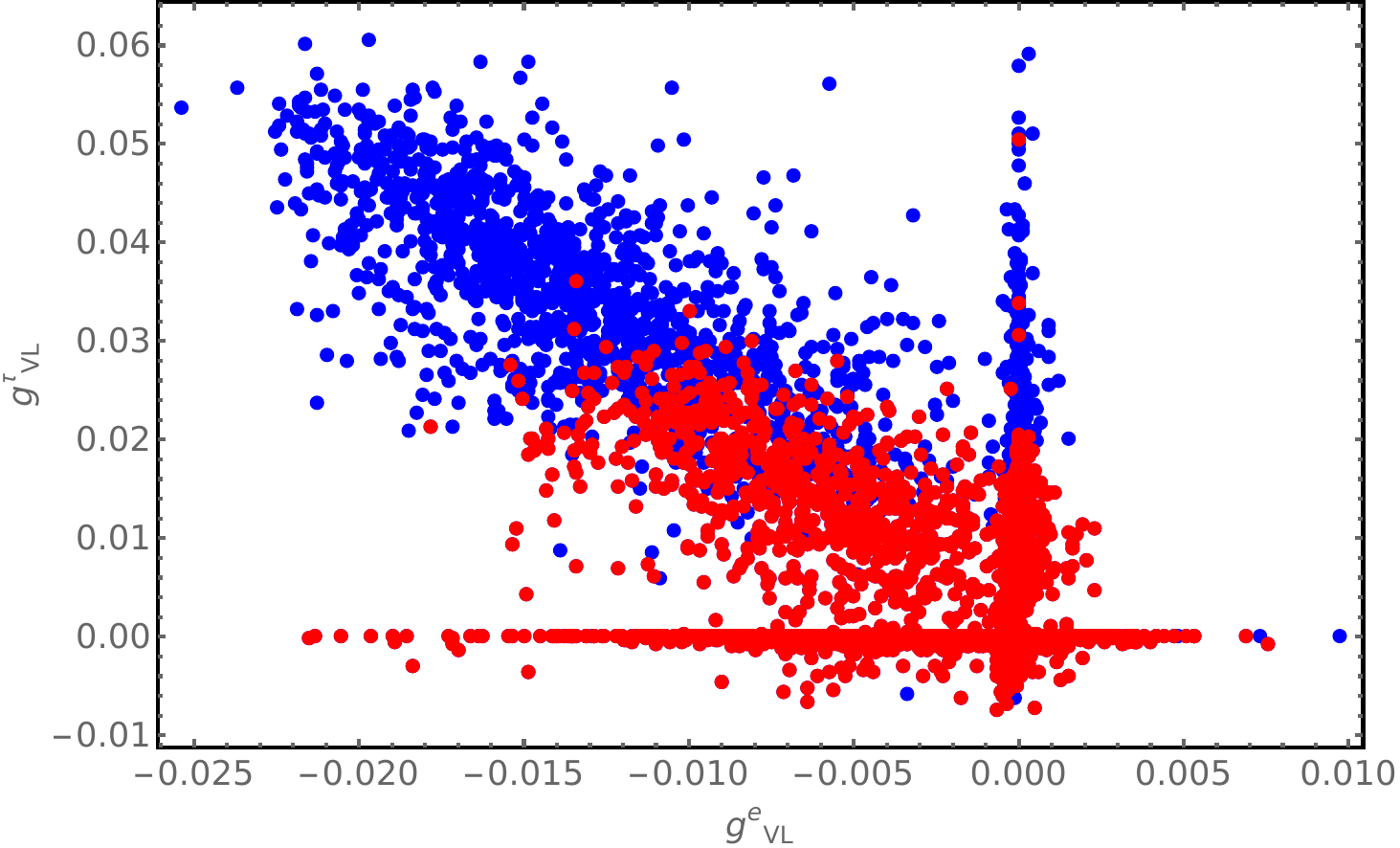}}
	\caption{Correlation between $g_{VL}^\tau$ and $g_{VL}^e$ in the BLSSM-IS. 
The different colors indicate regions of parameter space consistent with the SUSY mass bounds 
and flavor constraints.}
	\label{figgVL}
\end{figure*}

On the other hand, the right-handed sneutrino loops induce sizable corrections to the 
effective vertices $\ell \tilde\chi^- \tilde\nu^R$ and $\tilde\chi^0 \nu_\ell \tilde\nu^R$. 
These loop effects are governed by the large neutrino Yukawa couplings in the inverse seesaw 
mechanism and are generally flavor dependent. As illustrated in Fig.~\ref{figgVL}, the 
light-lepton couplings $g_{VL}^{e,\mu}$ can be shifted to negative values, while 
$g_{VL}^\tau$ remains positive. This asymmetry arises because the SM tree-level 
contribution dominates in the $\tau$ channel, whereas in the $e$ and $\mu$ channels the 
loop-induced terms can compete with and even outweigh the SM contribution, thereby driving 
$g_{VL}^{e,\mu}$ below zero.

A negative $g_{VL}^{e,\mu}$ reduces the decay rates $\Gamma_{e,\mu}$ relative to their SM 
values, while leaving $\Gamma_\tau$ essentially unaffected. Since the lepton-flavor 
universality ratios are defined as $\mathcal{R}(D^{(*)}) = \Gamma_\tau/\Gamma_{e,\mu}$, 
such a suppression of $\Gamma_{e,\mu}$ leads to a modest enhancement of 
$\mathcal{R}(D^{(*)})$. Through the sum rule correlation, the same effect propagates to 
$\mathcal{R}(\Lambda_c)$, resulting in a slight upward shift with respect to the SM 
prediction. This feature is clearly visible in Fig.~\ref{figgVL}, where the parameter space 
compatible with current experimental constraints allows for small but significant 
enhancements of both mesonic and baryonic ratios.

\section{Numerical Analysis and Results}

We now display our results for $\mathcal{R}(\Lambda_c)$ alongside with $\mathcal{R}(D)$ and $\mathcal{R}(D^{*})$. In our calculation and numerical analysis we used the SPheno codes \cite{Porod:2011nf} generated by SARAH \cite{Staub:2013tta}, to scan the parameter space of the low scale BLSSM-IS.

\begin{figure*}[h!]
	\centering
\includegraphics[width=15cm, height=6cm]{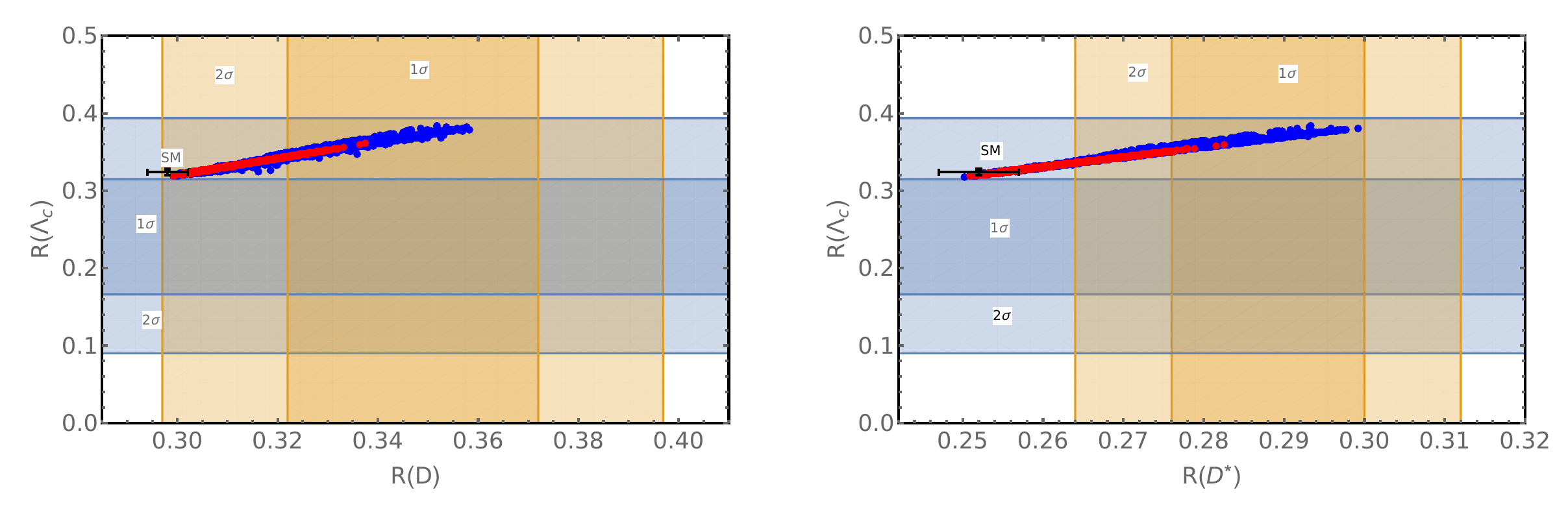}~
	\caption{Correlation between $\mathcal{R}(\Lambda_c)$ and 
$\mathcal{R}(D^{(*)})$ in the BLSSM-IS, with the same color coding as Fig.~\ref{figgVL}. 
The horizontal and vertical bands show the $1\sigma$ and $2\sigma$ experimental ranges, 
while the error bars indicate the SM predictions.
}
\label{fig5}
\end{figure*}

Fig.\ref{fig5} illustrates the interplay between the baryonic observable $\mathcal{R}(\Lambda_c)$
and the mesonic ratios $\mathcal{R}(D)$ and $\mathcal{R}(D^*)$. The figure shows the regions allowed 
by experimental data at the $1\sigma$ and $2\sigma$ levels, together with the 
SM expectations and the parameter points obtained within the 
BLSSM-IS framework. 

A key observation is that although the current central value of  $\mathcal{R}(\Lambda_c)$ 
reported by LHCb is somewhat lower than the SM prediction, its large experimental 
uncertainty (roughly $30\%$) implies that the SM still lies comfortably within 
its $2\sigma$ range. At the same time, the mesonic observables show a complementary 
trend: while both $\mathcal{R}(D)$ and $\mathcal{R}(D^*)$ prefer values above the SM predictions, 
their world averages remain consistent with the SM at the $2\sigma$ level. 
In fact, $\mathcal{R}(D)$  deviates by about $1.4\sigma$ and $\mathcal{R}(D^*)$  by about $2.5\sigma$. 

What is striking in Fig.~\ref{fig5} is the \emph{consistency band} imposed by the sum rule
\begin{equation}
\frac{\mathcal{R}(\Lambda_c)}{\mathcal{R}(\Lambda_c)_{\rm SM}}
= 0.28 \,\frac{\mathcal{R}(D)}{\mathcal{R}(D)_{\rm SM}}
+ 0.72 \,\frac{\mathcal{R}(D^*)}{\mathcal{R}(D^*)_{\rm SM}}
+ \delta_{\Lambda_c},
\end{equation}
where $\delta_{\Lambda_c}$ is small unless large scalar or tensor effects are present. 
This correlation implies that any upward shift in $\mathcal{R}(D)$ and $\mathcal{R}(D^*)$ (as suggested 
by experiment) necessarily drives $\mathcal{R}(\Lambda_c)$ to larger values compared to its SM 
prediction. Therefore, if $\mathcal{R}(D)$ and $\mathcal{R}(D^*)$ are both taken within their $1\sigma$ 
experimental ranges, the sum rule predicts $\mathcal{R}(\Lambda_c)$ values that lie above the 
SM line, close to or within the experimentally allowed $2\sigma$ band. 

In summary, Fig.\ref{fig5} highlights that the experimental values of $\mathcal{R}(\Lambda_c)$ at the 
$2\sigma$ level can be compatible with $\mathcal{R}(D)$ and $\mathcal{R}(D^*)$ taken within their $1\sigma$ 
ranges. While the SM predictions remain inside the $2\sigma$ experimental intervals, they 
do not fully account for the upward pull in $\mathcal{R}(D)$ and $\mathcal{R}(D^*)$ together with the lower 
central value of $\mathcal{R}(\Lambda_c)$. The sum rule correlation among these observables 
naturally drives $\mathcal{R}(\Lambda_c)$ to higher values once the mesonic ratios are fixed near 
their preferred $1\sigma$ regions, thereby improving the overall consistency and offering a 
powerful test of the effective Hamiltonian structure in future measurements. 

However, if the central value of $\mathcal{R}(\Lambda_c)$ remains as low as it is currently 
measured, a conflict with the sum-rule correlation would emerge. In this case, the only 
consistent explanation would be the presence of new-physics effects that generate a sizable 
contribution to $\delta_{\Lambda_c}$. Since $\delta_{\Lambda_c}$ receives its dominant 
enhancements from tensor operators, this scenario would strongly suggest non-negligible 
tensor interactions in the $b \to c \tau \nu$ transition. Such contributions would modify 
the correlation by lowering $\mathcal{R}(\Lambda_c)$ without simultaneously reducing $\mathcal{R}(D)$ and 
$\mathcal{R}(D^*)$, thereby reconciling the observed pattern. 

Alternatively, one might consider more exotic possibilities: a selective new-physics 
contribution to the baryonic channel that is suppressed or absent in the mesonic modes, 
perhaps arising from hadronic structure effects beyond the current lattice calculations or 
from operators with different interference patterns in baryonic versus mesonic decays. 
Nonetheless, these explanations are more constrained and less natural, since the effective 
Hamiltonian structure generically links mesonic and baryonic observables. Therefore, a 
persistently low value of $\mathcal{R}(\Lambda_c)$ compared to the correlated expectations from 
$\mathcal{R}(D)$ and $\mathcal{R}(D^*)$ would be a strong hint of tensor-like new physics, making future 
improvements in the measurement of  $\mathcal{R}(\Lambda_c)$ particularly decisive. 

\section{Conclusions}

In this work we have investigated the lepton flavor universality ratios 
$\mathcal{R}(D)$, $\mathcal{R}(D^{\ast})$, and $\mathcal{R}(\Lambda_c)$ within the 
$B-L$ Supersymmetric Standard Model with an inverse seesaw (BLSSM-IS). 
We have shown that penguin-type loop corrections involving charginos, 
neutralinos, and right-handed sneutrinos can induce flavor-dependent 
shifts in the effective couplings $g_{VL}^\ell$. In particular, 
$g_{VL}^{e,\mu}$ can receive negative loop contributions, while 
$g_{VL}^\tau$ remains positive, leading to a suppression of 
$\Gamma_{e,\mu}$ relative to $\Gamma_\tau$. As a result, 
$\mathcal{R}(D^{(\ast)})$ is modestly enhanced compared to its Standard 
Model prediction, and through the sum-rule relation, 
$\mathcal{R}(\Lambda_c)$ also receives a small upward shift. 

Our numerical analysis demonstrates that these effects remain consistent 
with present constraints from flavor observables, electroweak precision 
data, and direct SUSY searches. Although the current experimental 
uncertainties in $\mathcal{R}(\Lambda_c)$ are still large, our results 
highlight a potential correlation with the mesonic ratios that can be 
tested more stringently in future measurements. In particular, improved 
precision at Belle II and the LHCb upgrade will be decisive in probing 
whether the modest enhancements predicted by the BLSSM-IS framework are 
realized in nature.

\section*{Acknowledgments}
DB is supported by the Algerian Ministry of Higher Education and Scientific Research
under the PRFU Project No. B00L02UN400120230002. 
SK is partially supported by the Science, Technology and Innovation Funding Authority (STDF) under
Grant No. 48173. 
\bibliographystyle{JHEP}
\bibliography{RDbib}

\end{document}